\documentstyle[aps,prl,graphicx,floats]{revtex}

\begin{document}
\draft
\twocolumn[\hsize\textwidth\columnwidth\hsize\csname @twocolumnfalse\endcsname

\title{ Josephson transport through a Hubbard impurity center.}

\author{V.I. Kozub, A.V. Lopatin and  V.M. Vinokur}
\address{
Material Science Division, Argonne National Laboratory,
Argonne, Illinois 60439 \\
A.F. Ioffe Physico-technical Institute, 194021, St.-Petersburg,
Russia }
\date{\today}
\maketitle

\begin{abstract}
We investigate the Josephson transport through a thin semiconductor barrier
containing impurity centers with the on-site Hubbard interaction $u$ of an 
arbitrary sign and strength. We find that in
the case of the repulsive interaction the Josephson current
changes sign with the temperature increase if the energy of the
impurity level $\varepsilon$ (measured from the Fermi energy of
superconductors) falls in the interval $(-u,0)$. We predict strong
temporal fluctuations of the current if only a few centers present
within the junction. In the case of the attractive impurity
potential ($u<0$) and at low temperatures, the model is reduced to
the effective two level Hamiltonian allowing thus a simple
description of the nonstationary Josephson effect in terms of pair
tunneling processes.

  \end{abstract}

\vskip2pc]

\bigskip

  Josephson transport through a semiconductor barrier containing
resonance impurity centers is a subject of  intense current
theoretical and experimental interest. 
Being a coherent current of Cooper pairs, the Josephson current 
flowing through the impurity center is extremely sensitive
to the presence of on-site impurity interaction
\cite{Matveev_Glazman,Kivelson}. This offers a  
possibility of using the Josephson effect as a unique spectroscopic tool for 
measurements of impurities energy states 
and calls for a theoretical study  of Josephson transport through an
impurity level with arbitrary strength and sign of the on-site interaction.

The on-site Coulomb repulsion makes the occupation of an impurity
center by a Cooper pair unfavorable.  Thus, one would expect that
the Josephson current flowing through an impurity level is
considerably suppressed (unlike the normal current via a resonance
state).  However, recent analysis of the hopping magnetoresistance
data for different semiconductors, revealed a presence of a
comfortable transmission channel via the double occupied level
(related to the states of the upper Hubbard band) \cite{ours} with
the very low, as compared to the naively expected value, repulsion
energy.  Such a repulsion reduction may result from the polaronic
effect, which sometimes can even ``overscreen" the Coulomb
repulsion reverting it to the effective {\it attraction} at the
site.  The so-called D$^X$-centers in semiconductors formed by
substitutional dopants in $GaAs$ and $AlGaAs$ alloys (see
e.g.\cite{DX1,DX2}) represent an example of those attractive
impurities.

In this Letter we investigate the Josephson transport through a
semiconductor barrier with impurity centers of arbitrary 
on-site interaction strength and sign.  
We restrict ourselves to a sufficiently low impurity concentration 
such that the presence
of an electron on a given site does not affect its neighbors. We
consider the semiconductor film to be thin enough, so that it's width
is less than the average distance between the impurity centers: in
this case the transport is indeed determined by the tunneling through
a single impurity rather than by hopping over a chain of resonant
impurities levels.

Given an on-site interaction strength, $u$, the occupancy of the
impurity level is controlled by the level energy $\varepsilon.$ For
$u>0$ (repulsion),  the impurity level is double occupied if
$\varepsilon<-u$. It is single-occupied when $-u <\varepsilon<0$,
and, finally, the impurity state is empty if $\varepsilon>0$. In case
of very strong repulsion $u \to \infty$ and zero temperature, 
the Josephson current changes sign abruptly as soon as the transition 
from the empty state to a single-occupied state occurs,
in the later case a so-called $\pi$-junction is realized  
\cite{Matveev_Glazman}. 
We show that this effect holds and that the jump of the Josephson current
becomes even more pronounced for any finite positive interaction
$u$. At finite temperatures in the regime $-u<\varepsilon<0$ 
the double - and zero occupied states that carry positive Josephson
currents become exited. We demonstrate that it results 
in nonmonotonic dependence of the Josephson
current on temperature: the current first increases from 
negative to positive values and then decreases.

In case of an attracting center, the impurity level is either
double occupied, if $\varepsilon<u/2$, or empty  when $\varepsilon
> u/2$. Close to the resonance, when $\varepsilon \approx u/2 $, 
the main contribution to the
Josephson current comes from pair tunneling processes, and the
system can be described by the effective two level Hamiltonian
that includes pair tunneling processes only
\cite{Kivelson}. Solution of this model gives two energy
branches with energies $E_{\pm}(\phi)$ that depend on the phase
difference $\phi=\phi_2-\phi_1$ of the superconductors and
correspond to exactly opposite Josephson currents (see Fig.
\ref{energy_br}).
 At finite temperatures the upper level is exited and the Josephson 
current depends on the
temperature as $\tanh E(\phi)/2T$ with
$E(\phi)=E_+(\phi)-E_-(\phi)$. This model also allows for a simple
enough analysis of the nonstationary Josephson effect in the
regime where the applied voltage $ V \ll min(\Delta, u)/e$.
We will consider the simplest set-up where the Josephson current
is shunted by the resistor and show that in this regime the
current-voltage dependence exhibits resonant peaks.

{\it Model.} We describe our system by the Hamiltonian of impurity level 
coupled with  superconductors via weak tunneling matrix elements 
 $t_{1}$ and $t_2$ 
$$
\hat H=\hat H_1+\hat H_2+\hat H_0 + 
\sum_{k=1,2} t_k\, [\hat \psi_{k\alpha}^\dagger(0)\, \hat d_\alpha+
\hat d_\alpha^\dagger \psi_{k\alpha}(0)]  
$$
where $H_1$ and $H_2$ are the BCS Hamiltonians of the
superconductors and $H_0$ is the impurity Hamiltonian $ \hat
H_0=\varepsilon \hat d^\dagger_\alpha\, \hat d_\alpha  + u\, \hat
d^\dagger_\uparrow \hat d_\uparrow\, \hat d^\dagger_\downarrow
\hat d_\downarrow. $

  {\it  Perturbation theory in tunneling matrix elements.}
The current flowing through the impurity can be found as
the expectation value of the  current operator
$
\hat I_1=t_1\, 
e\,i\,[\hat \psi_{1\alpha}^\dagger(0)\,\hat d_\alpha-\hat d_\alpha^\dagger
 \hat \psi_{1\alpha}(0)]$
flowing between the superconductor 1 and the impurity.
In the second order perturbation theory in $t_1$ the 
current $I=\langle I_1 \rangle $ is
\begin{equation}
I=-4 i\,e\,t_1^2\, {\rm Im}\,
 \int_0^\beta\, d\tau_1 \, F_{1\psi}^\dagger(\tau)\, F_d(-\tau),  \label{I_1}
\end{equation}
 where $F_{1\psi}(\tau)$ 
is the local anomalous Green function of the superconductor 1,
$ F_{1\psi}(\omega)= \pi\, \nu_1\,\Delta_1 /
\sqrt{|\Delta_1|^2+\omega^2 } $, where $\nu_1$ is the density of
states. The anomalous Green function of the impurity center
$F_d(\tau_1,\tau_2)= \langle d_\uparrow(\tau_1)
d_\downarrow(\tau_2) \rangle$ is found by the second order
perturbation theory in $t_2$ and the current  $I$ becomes
 \begin{eqnarray}
I=-4\,i\,e\,t_1^2\,t_2^2\int d \tau \, d\tau_3 && \, d\tau_4
\,F_{1\psi}^\dagger(\tau)     \nonumber    \\
&&\times K(0,\tau,\tau_3,\tau_4)\, F_2(\tau_4-\tau_3),
\end{eqnarray}
where the correlation function
\begin{equation}
K(\tau_1,\tau_2,\tau_3,\tau_4)=
\langle d_\uparrow(\tau_1)\, d_\downarrow(\tau_2)\,
d^\dagger_\downarrow(\tau_3)\, d^\dagger_\uparrow (\tau_4) \rangle_d,
\end{equation}
is defined with respect to impurity Hamiltonian $H_0.$ Assuming
for simplicity that $|\Delta_1|=|\Delta_2|=\Delta$ we arrive at the
final answer for the current in the form
\begin{equation}\label{1}
I=I_0(\phi) \left(i_0+i_1\,e^{-\beta\varepsilon}+i_2
\,e^{-\beta\varepsilon_2}    \right)\,Z^{-1} \label{Answer}
\end{equation}
where $\varepsilon_2=2\varepsilon+u$ is the energy of double-ocupied 
state, 
\begin{equation}\label{2}
I_0(\phi)=4\,\pi^2\,e\, (t_1\,t_2)^2 \,\nu_1\, \nu_2\, \sin(\phi),
\end{equation}
the impurity partition function
$Z=1+2e^{-\beta\varepsilon}+e^{-\beta\varepsilon_2},$ and the terms
$i_0$ and $i_2$ are
\begin{eqnarray}
i_0&=&{1\over{\varepsilon_2 }}A^2(\varepsilon)+
{1\over{\Delta}} B(\varepsilon),  \nonumber    \\
i_2&=&{-1\over{\varepsilon_2 }}A^2(\varepsilon-\varepsilon_2)
+{1\over{\Delta}}
B(\varepsilon-\varepsilon_2),    \nonumber
\end{eqnarray}
with functions $A$ and $B$ defined by
\begin{eqnarray}
\label{A_definition}
A(\epsilon)&=&2\,T \sum_\omega \, {{\epsilon }\over{\epsilon^2+\omega^2 }}\,\,
{{\Delta }\over{\sqrt{\Delta^2+\omega^2} }}, \\
B(\epsilon)&=&- \Delta
\,A^2(\epsilon)/2\epsilon+T\sum_\omega {1 \over {\epsilon^2+\omega^2 }}\,\,
{{ \Delta }\over{ \Delta ^2+\omega^2 }}. \nonumber
\end{eqnarray}
\hspace{0.5cm}
 \begin{figure}[ht]
\includegraphics[width=3.3in]{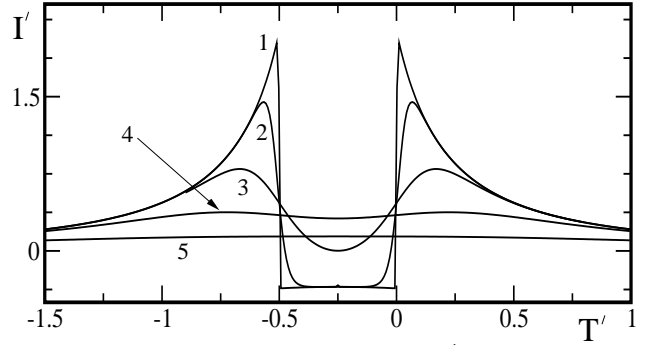}
 \caption{The Josephson current
$I^\prime=I/I_0(\phi)$
as a function of the energy of the single-occupied impurity state
$\varepsilon$ for $u/ \Delta=0.5$ and different temperatures:
$T^\prime=T/\Delta=0,0.02,0.04,0.1,0.2$ for plots 1-5 respectively.  }
\label{repulsive}
\end{figure}
The contribution  $i_1$  in (\ref{Answer}) is
\begin{eqnarray}
i_1={{A^2(\varepsilon)-A^2(\varepsilon-\varepsilon_2)} \over{\varepsilon_2}}
-{{B(-\varepsilon)+B(\varepsilon_2-\varepsilon)) }\over{\Delta}}   \nonumber  \\
+4\,T\sum_\omega\, {{\varepsilon(\varepsilon_2-\varepsilon)+\omega^2 }
\over{(\varepsilon^2+\omega^2)((\varepsilon_2-\varepsilon)^2+\omega^2 )}}\,\,
 {{\Delta^2 }\over{\Delta^2+\omega^2 }}.
\end{eqnarray}

{\it Repulsive Hubbard center. } In the case of repulsive
interaction at zero temperature Eq.(\ref{Answer}) simplifies to
\begin{equation}  \label{j_zero}
I=I_0(\phi)\times \left \{
\begin{array}{ccc}
i_2,&  \;\;\;\;\;\;\; \varepsilon <-u \\
i_1/2,&  \;\;\;\;\;\;\;\;\; -u< \varepsilon <0 \\
i_0, &  \;\;\;\;\;\;\; \varepsilon >0,
\end{array}
   \right.
\end{equation}
that describes double-occupied, single occupied and unoccupied
contributions respectively. The dependence of the Josephson
current on $\varepsilon$ given by Eq.(\ref{j_zero}) is
discontinuous (see Fig.1) with the jumps at points
$\varepsilon=0$ and $\varepsilon=-u$ 
where the Josephson current changes sign.
The maximal values of the Josephson current
$ I_{max}/I_0(\phi)=1/u +( 2/\pi  -1/2)/\Delta $
 grow as $1/u$ when $u\to 0$ such that for
$u<t^2\nu$ the perturbation theory  in hopping elements at
$\varepsilon\approx 0,-u$ becomes unapplicable. In the region
$-u<\varepsilon<0$ the Josephson current does not exibit a
singular behavior because ocupation of the impurity level by one 
electron prevents from the resonant
pair current flow even in case of small $u.$ Thus the height of
the jumps of the Josephson current at points $\varepsilon=0$
and $\varepsilon=-u$ decreases with the increase of the 
interaction strength $u$.

At finite temperatures the dependence of the Josephson current on
$\varepsilon$ becomes continuous  (Fig.1) and,
eventually, at high enough $T$ the Josephson current becomes
positive for all $\varepsilon.$ Thus for $\varepsilon$ laying
within the interval
 $(-u,0)$ the Josephson current changes it's sign as 
temperature grows (Fig.2a). This can be understood as follows:
At low temperatures the current is given by the single occupied
contribution $i_1$ which is negative, at finite temperature the
zero and double occupied states corresponding to positive currents
are exited. Since the latter ones make larger contributions, the
Josephson current eventually changes sign with the 
increase of temperature.

\vspace{0.7cm}
\begin{figure}[ht]
\hspace{0.5cm}
\includegraphics[width=3.3in]{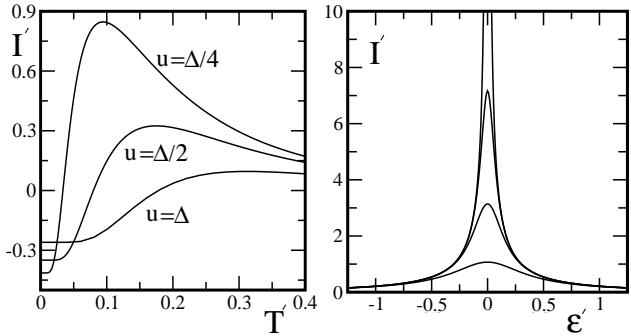}
 \caption{ a) The Josephson current
$I^\prime=I/I_0(\phi)$ as a function of temperature $T^\prime=T/\Delta$
for different positive inteactions $u$ and $\varepsilon=-u/2.$ b)
Dependence of the Josephson current on
$\varepsilon^\prime=\varepsilon+u/2$ 
for $u/ \Delta=-0.5$ and different temperatures:
$T^\prime=0,0.05,0.1,0.2$ (top to bottom).  }
\label{rep_temp_dep}
\end{figure}
\vspace{-0.1cm}

The temperature dependence in question implies the statistical
averaging over impurity centers. If only few centers are
present, one expects significant temporal fluctuations of the
critical current. These fluctuations can be observed, for example,
by including the Josephson junction in a superconducting
loop. The energy of such a loop in the abscence of external magnetic
field can be written as a function of magnetic flux $\Phi=LI,$
with $L$ being the inductance of the loop, as
$E_l(\Phi)=-I_c\cos(2\pi\Phi/\Phi_0)+\Phi^2/2 L,$
where $I_c$ is the critical current defined by
Eqs.(\ref{1},\ref{2}) and $\Phi_0$ is the magnetic flux quantum.
Assuming  $I_cL \gg \Phi_0$ 
we see that for positive $I_c$ the minimal solution corresponds to $\Phi=0$
while for $I_c < 0$ the solution corresponds to $\Phi
= \Phi_0/2$. Transitions between different states of the impurity
center lead to fluctuations of the magnetic flux through the loop.
Change in the occupation number requires a quasiparticle tunneling
from the leads to the center or vice versa. Thus, if $T < \Delta$
the characteristic fluctuation time is of the order of $\tau^{-1}
\sim (t^2/\hbar) \nu \exp(- \Delta/T)$. The characteristic times
spent in the empty, single-occupied and double-occupied states are
given as $\tau Z$, $\tau Z \exp( \beta\epsilon)$ and $\tau Z \exp(
\beta \epsilon_2)$, respectively.

{\it Attractive Hubbard center.} At $u<0$ the dependence of the
Josephson current on the level energy $\varepsilon $ shows a
resonance at values $\varepsilon\approx|u|/2,$ where the energy of
the double-occupied state $\varepsilon_2$ approaches zero (see Fig.2b).
 Near the resonance Eq.(\ref{Answer}) can be simplified leaving only the 
terms that demonstrate singular behaviors $1/\varepsilon_2 $
$$
I={{I_0(\phi)}\over{\varepsilon_2}}
{{
A^2((\varepsilon_2+|u|)/2)
-A^2((\varepsilon_2-|u|)/2)\, e^{-\beta\varepsilon_2} }\over{
1+ e^{-\beta\varepsilon_2}}}
$$
 For $T\ll\Delta$ the maximal value of the current is
\begin{equation}
I_{max}=I_0(\phi)\, A^2(|u|/2)/2T. \label{res_val}
\end{equation}
 This expression diverges at very low temperatures $T <t^2\nu$ where the
 approach based on the direct perturbation theory in tunneling elements
 becomes unapplicable.

{\it Effective low temperature model.} In case of the
attractive interaction the  Josephson current obtained by the
perturbation theory in $t_1$ and $t_2$ has a form of the Gibbs average
of two terms corresponding to unoccupied and double occupied
states. This holds as long as $T\ll u$, and the system in this
regime can be described by the effective Hamiltonian 
\cite{Kivelson,Matveev}
$$
\hat H_{eff}=\varepsilon_2 \hat b^\dagger \hat b
+\tilde t_1\,[\hat b^\dagger e^{i\phi_1}+\hat b e^{-i\phi_1}] +
\tilde t_2\, [\hat b^\dagger e^{i\phi_2}
+\hat b e^{-i\phi_2}],
% \label{effect_model}
$$
 where $\hat b =\hat d_\downarrow \, \hat d_\uparrow$  is the hard-core boson operator
satisfying $ [\hat b,\hat b^\dagger]=1-2\hat b^\dagger\hat b, \;\;
\hat  b^\dagger\,\hat b^\dagger=0. $ The eigenfunction of the
Hamiltonian $ \hat H_{eff}$ can be written  as
$
\psi=\alpha |0> +\beta |1>,
$
where $|0>$ is the empty state and $|1>$ is the double occupied
(one pair) state.
The
operator of the current flowing through the impurity is
\begin{equation}
\hat I=2e\,i\,\tilde t_1 <\psi|\, \hat b\, e^{-i\phi_1}- b^\dagger\, 
e^{i\phi_1} |\psi>.
\end{equation}
Solving the Schrodinger equation $ \hat H_{eff}\, \psi=E\, \psi
\label{schr_eq} $ we find two eigenstates with  energies
\begin{equation}
\label{energy_eff}
E_{\pm}(\phi)=\varepsilon_2/2\pm \sqrt{ \delta^2+E_0^2\,(\cos(\phi)+1)}
\end{equation}
where $\delta^2=\varepsilon_2^2/4+(\tilde t_1
-\tilde t_2)^2,\,$ $E_0^2=2 \tilde t_1 \tilde t_2$. The currents
corresponding to these two states are
\begin{equation}
\label{current_eff}
I_{\pm}(\phi)=\mp {{e E_0^2\,\sin(\phi) }\over {
\sqrt{ \delta^2+ E_0^2(\cos(\phi)+1)  } }}= 
2e {{d\, E_{\pm}(\phi) }\over{d\,\phi}}.
\end{equation}

  Dependences of the energies $E_{\pm}$ and currents $I_{\pm}$ on
the phase $\phi$ are shown on Fig 3.
Comparing the current $I_{-}$ given by Eq.(\ref{current_eff})
 with the result of  perturbation theory in $t_1,t_2$
we find the effective tunneling matrix elements
$
\tilde t_{1,2}=\pi t_{1,2}^2 \nu_{1,2}\, A(|u|/2).
$
 At $T\ll u$ the summation in
(\ref{A_definition}) can be reduced to the integration resulting in 
\begin{equation}
A(\epsilon)= \left\{
\begin{array} {cc}
{{1-(2/\pi)
\,\arcsin(\epsilon / \Delta\,)}\over{
\sqrt{1-\epsilon^2/ \Delta^2 }}} \;\;\;\;\;\;\;\;\;\;\;\;\;\;\;
0<\epsilon<\Delta    \\
{{ \Delta  }\over{\pi\,\sqrt{\epsilon^2-\Delta^2 }}}
\; \ln {{\epsilon+\sqrt{\epsilon^2- \Delta^2} }\over{\epsilon-
\sqrt{\epsilon^2-\Delta^2}
}}\;\;\;\;
\epsilon > \Delta.
\end{array}
     \right.  
\end{equation}
At finite temperatures the Josephson current is given by the
thermodynamic average of the two states $ I(\phi)=I_-(\phi)\,
\tanh E(\phi)/2T $ with $E(\phi)=E_+(\phi)-E_-(\phi)$.  As in case
of repulsive interaction, for the junction containing only few
centers we expect strong temporal fluctuations of the current
since the critical currents for two branches have opposite signs.
In particular, for the weakly coupled superconducting loop
controlled by a single center within the Josephson junction the
flux fluctuations between the values $\Phi = 0$ and $\Phi =
\Phi_0/2$ are expected according to the discussion given above.

\begin{figure}[ht]
\hspace{0.5cm}
\includegraphics[width=3.3in]{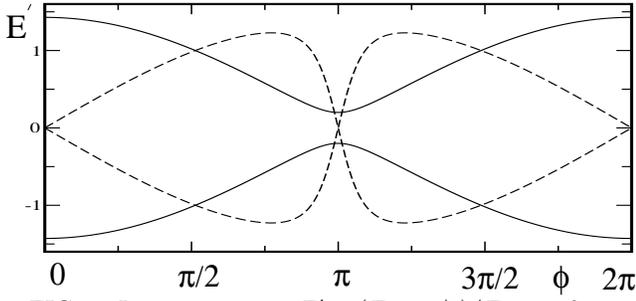}
 \caption{ Impurity energy
 $E^\prime=(E-\varepsilon_2/2)/E_0$ as a function the  phase
$\phi=\phi_1-\phi_2$ (black lines)
and the corresponding Josephson
currents $I_{\pm}^\prime=I_{\pm}/(eE_0)$ (dashed lines).}
\label{energy_br}
\end{figure}

 {\it Nonstationary regime.} Let us consider the simplest experimental
set-up (see Fig 4.):
 Superconductor
contacts are shunted by the resistor 
while the potential of the superconductor 1 and the current $I_0$
are  fixed. The potential of the superconductor 2 $V_2$ is determined by
current conservation equation
\begin{equation}
V_2/R+I_2(t)=I_0,  \label{sys1}
\end{equation}
where $I_2(t)$ is the current flowing between the impurity center
and the superconductor 2
\begin{equation}
I_2(t)=-2ei\, t_2 \, Im\{\alpha^*  \beta \, e^{-i\phi_2}  \}
\label{sys2}
\end{equation}
The phase $\phi_2$ is related  to the voltage $V_2$ by
\begin{equation}
\dot \phi_2=-2 e V_2   \label{sys3}.
\end{equation}

 To close the system of equations (\ref{sys1}-\ref{sys3})
 we write the Schrodinger equation for the amplitudes $\alpha,\beta$
 \begin{eqnarray}
 i\dot \alpha = \beta \,  \tilde t_1  +\beta\,  \tilde t_2\, e^{-i\phi_2}  \\
 i\dot\beta =\beta \, \epsilon_i + \alpha \, \tilde t_1
+\alpha \,  \tilde t_2\,  e^{i\phi_2}.
\label{shr_eq}
 \end{eqnarray}
The Cooper pair energy on the impurity site $\varepsilon_i$ has a contribution
from the potential of the superconductor 2 
$\varepsilon_i=\varepsilon_2+2ke V_2$ where 
$k=l_1/(l_1+l_2)$ and $l_1,l_2$ are the distances between the impurity center
and superconductors 1 and 2 respectively. Eqs.(\ref{sys1}-\ref{shr_eq})
can be easily solved numerically. Solution for $k=0,
\epsilon_2/E_0=0.5, \tilde t_1/E_0=2.0,
\tilde t_2/E_0=0.5$ and different shunting resistors
are shown on Fig. 4 along with approximate adiabatic solutions
shown by the dashed lines. The adiabatic
approximation is valid in the low voltage regime where
the phase $\phi_2$ changes slowly and the current is 
given by
\begin{equation}
\label{current_nonst} I_{2}(t)=-e E_0^2 \sin({\phi_2} t)/ \sqrt{
\delta^2+ E_0^2(\cos({\phi_2}t)+1) },
\end{equation}
\begin{figure}[ht]
\hspace{0.5cm}
\includegraphics[width=3.2in]{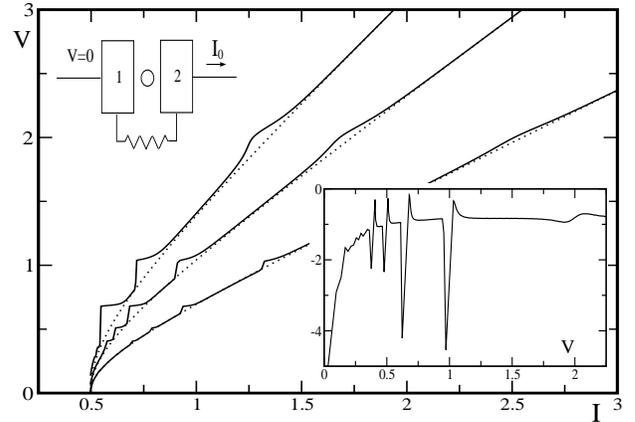}
\caption{
Voltage-current dependence of the shunted Josephson contact
for receptivities $R e^2 /\hbar=1.6,1.2,0.8$ (top to bottom). 
Upper inset shows the setup while the lower one shows
the differential resistance ($dV/dI$ ) as a function of the applied Voltage
corresponding to the lower graph ($R e^2 /\hbar=0.8$). }
\label{IV}
\end{figure}
corresponding to the lower energy branch $E_{-}(\phi_2)$. At
higher voltages the effect of Landau-Zenner tunneling between two
energy branches cannot be neglected. As we see it from the
numerical solution, transitions between the branches appear as
resonances on the current-voltage dependence. Positions of these
resonances can be easily estimated in case $t_1 \gg  t_2$; taking
$t_2=0$ in Eq.(\ref{energy_eff}) we obtain that the energy levels
are splitted by $\Delta E=2\sqrt{\varepsilon_2^2/4+\tilde t_1^2}$.
Resonances arise when the  frequency of the phase oscillations of
the superconductor 2 $\omega=2e V_2/\hbar$ is related with 
$\Delta E$ as $m\hbar\,\omega\, =\Delta E$.  Shown in the the inset of
Fig. 4 is the differential resistance as a function of voltage.
One clearly sees resonances satisfying the above conditions.

{\it Conclusions.} In conclusion we considered the Josephson
transport through an impurity center for the cases of arbitrary
sign and arbitrary strength of the Hubbard interaction. For
repulsive centers we show that the Josephson current changes sign
with temperature when the energy of single occupied states
$\varepsilon$ lays within the interval $(-u,0).$ If the junction
contains only few centers strong temporal fluctuations of the
current are predicted. In the case of attractive centers we study
the nonstationary Josephson effect with the help of the effective
model that takes into account only pair tunneling processes. We
consider the case of resistivelly shunted Josephson junction and
show that current-voltage characteristic has resonances associated
with transition between two states formed due to coupling of the
impurity with one of the superconductors.

We would like to thank Y.M. Galperin, A.E. Koshelev and K. Matveev for
useful discussions. This work was supported by the 
U.S. Department of Energy, Office of Science under contract 
No. W-31-109-ENG-38.

\end{document}